\documentclass[a4paper,12pt]{article}
\usepackage{graphicx}
\usepackage[affil-it]{authblk}
\usepackage[top=1in, bottom=1in, left=1in, right=1in]{geometry}
\usepackage{dcolumn}
\usepackage{bm}
\usepackage{color}
\usepackage{amssymb}
\usepackage{amsmath}
 \usepackage[square, comma, sort&compress, numbers]{natbib}
\usepackage{multirow,setspace,times,amssymb,amsmath,graphicx,color,rotating,subfigure,url}

\begin{document}

\title{Mutual Feedback Between Epidemic Spreading and Information Diffusion}

\author{
Xiu-Xiu Zhan{\small$^{\mbox{1,2}}$},
Chuang Liu{\small$^{\mbox{1}}$},
Ge Zhou{\small$^{\mbox{1}}$},
Zi-Ke Zhang{\small$^{\mbox{1}}$},
Gui-Quan Sun{\small$^{\mbox{3}}$}
\& Jonathan J. H. Zhu{\small$^{\mbox{4}}$}}

\affil{
$^{1}$Alibaba Research Center for Complexity Sciences, Hangzhou Normal University, Hangzhou, 311121, P. R. China\\
$^{2}$Department of Mathematics, North University of China, Taiyuan, Shan'xi 030051, P. R. China\\
$^{3}$Complex Sciences Center, Shanxi University, Taiyuan 030006, P. R. China\\
$^{4}$Web Mining Lab, Department of Media and Communication, City University of Hong Kong, Kowloon, Hong Kong Special Administrative Region
}

\maketitle

\begin{abstract}
The impact that information diffusion has on epidemic spreading has recently attracted much attention. As a disease begins to spread in the population, information about the disease is transmitted to others, which in turn has an effect on the spread of disease. In this paper, using empirical results of the propagation of H7N9 and information about the disease, we clearly show that the spreading dynamics of the two-types of processes influence each other. We build a mathematical model in which both types of spreading dynamics are described using the SIS process in order to illustrate the influence of information diffusion on epidemic spreading. Both the simulation results and the pairwise analysis reveal that information diffusion can increase the threshold of an epidemic outbreak, decrease the final fraction of infected individuals and significantly decrease the rate at which the epidemic propagates. Additionally, we find that the multi-outbreak phenomena of epidemic spreading, along with the impact of information diffusion, is consistent with the empirical results. These findings highlight the requirement to maintain social awareness of diseases even when the epidemics seem to be under control in order to prevent a subsequent outbreak. These results may shed light on the in-depth understanding of the interplay between the dynamics of epidemic spreading and information diffusion.
\end{abstract}

\linespread{2}

\newpage
\newcommand{\rmnum}[1]{\romannumeral #1}
\newcommand{\Rmnum}[1]{\expandafter\@slowromancap\romannumeral #1@}
\makeatother
\section{Introduction}
Understanding how diseases spread among individuals has been a recent area of intense investigation ~\cite{Pastor-Satorras2014}. Epidemic spreading is generally considered to be a dynamic process in which a disease is transmitted from one individual to another through contacts in peer-to-peer networks. To date, a vast amount of research has focused on understanding the epidemic spreading phenomenon, which can be categorized into the following types: (1) epidemic spreading dynamics in various network structures \cite{Durrett2010}, such as a scale-free network \cite{Pastor-Satorras2001,Barthelemy2004}, a small-world network \cite{Kuperman2001,Kleczkowski2012} and an interdependent network \cite{Dickison2012,LiDQ2014}; (2) propagation mechanisms that describe the epidemic spreading process, such as the Susceptible-Infected-Recovered (SIR) model for influenza ~\cite{Hufnagel2004,Germann2006}, the Susceptible-Infected-Susceptible (SIS) model for sexually transmitted disease ~\cite{Gomez-Gardenes2008,Risau-Gusman2012} and the Susceptible-Exposed-Infected-Recovered (SEIR) model for rabies ~\cite{Childs2000,Zhang2011}; (3) data-driven modeling approaches that tackle the epidemic transmission ~\cite{Vespignani2012} by analyzing the available real datasets, such as the scaling laws in human mobility ~\cite{Vazquez2007,Meloni2011}, individual interactions ~\cite{Starnini2013,Karsai2011}, and contact patterns ~\cite{Holme2012,WangLin2013}.

The majority of the aforementioned studies focused on epidemic spreading independently, ignoring the fact that information diffusion of the diseases themselves also has a significant impact on epidemic outbreaks. For example, the outbreak of a contagious disease in a population leads to the spread of information, either through the media or friends, regarding the disease. This information impacts the protective measures that others may take such as staying at home, wearing face masks, and taking vaccinations \cite{Ruan2012}. These preventive behavioral responses upon receiving information regarding the disease may reduce the size of the epidemic outbreak. This is supported by research that has shown that people's behavioral responses contribute to the control of disease spreading, an example of which that is evident in the severe acute respiratory syndrome (SARS) in China in 2003 \cite{WorldHO2003}. Therefore, there has been an increased interest in examining the interaction between epidemic spreading and information diffusion. Seminal study was conducted in order to understand how the diffusion of awareness, or disease information, affects the spread of diseases. This was done by building a model in which the spread of both the disease and information in a host population was accounted for \cite{Funk2009,Funk2010a,Funk2010b,Funk2010c}. The results indicated that the interaction between these two different processes of spreading decreased the size of the epidemic outbreak in a well-mixed population \cite{Funk2009}. In some cases, enough behavioral changes would occur in response to the diffusion of disease information that the growing epidemic would vanish completely, this occurred even in cases where the epidemic transmission rate was higher than the classical threshold \cite{Sahneh2012,Wu2012}. Additionally, the interplay between information diffusion and epidemic spreading was elucidated on multiplex networks, where each type of spreading dynamics occurs on its own layer (information diffusion on communication layer versus epidemic spreading on physical layer) \cite{Cranell2013,Cranell2014,WangW2014}. The epidemic threshold as it relates to the physical contact layer can be increased by enhancing the diffusion rate of information on the communication layer. The effect that behavioral changes have on a population can be explained as being affected by the following three characteristics \cite{Funk2010b}: (i) the disease state of the individuals, such as vaccination \cite{Bauch2004,ZhangHF2014}; (ii) the epidemic transmission rate and recovery rate \cite{Cranell2013,Kiss2010}; (iii) the contact network structure, such as the adaptive process \cite{Gross2006,Fenichel2011,Liu2014}. Mass communication scholars share similar views on the causal linkages between the spread of epidemics and the diffusion of information regarding the epidemics.  For example, the outbreak of severe epidemics usually attracts heavy media coverage, subsequently resulting in the following three types of responses from the public: cognitive responses, such as the attention to the information and an increased awareness of the problem \cite{Zhu1992}, affective responses, such as anxiety, fear, or even panic \cite{Lupton2013}, and behavioral responses, such as the adoption of new practices in order to replace undesirable habits \cite{Rogers2003}.  However, these are more assumptions than empirical facts because it is difficult to find relevant data for such a clear-cut process and, even when the data are available, it is difficult to separate the unique effects of information on the control of epidemics from other confounding factors, such as changes in the diseases, the seasons and the medical treatments.

However, due to the difficulty in separating the unique effects of information on the control of epidemics from other confounding factors, such as changes in the diseases, the seasons, and the medical treatments, and the lack of relevant data,  these can be regarded as assumptions as opposed to empirical facts.

Recent studies regarding the interplay between information diffusion and epidemic spreading have focused on the suppression of epidemic outbreaks by information diffusion. The occurrence of a disease prompts the sharing of information regarding the disease, which leads to preventive measures that inhibit epidemic spreading \cite{Funk2009,WangW2014}. Studies have also indicated that when an outbreak is under control people are less vigilant in sharing relevant information which leads to a decrease in preventative measures and may result in a recurrence of pandemic epidemics. Unfortunately, evidence suggests that the second outbreak of epidemics is  typically more deadly. For example, the spread of SARS was shown to be alleviated in early March 2003, however, later that month showed a rapid increase, as indicated in the evolution curve of the probable cases of SARS (Fig. 2 in Ref. \cite{WorldHO2003}).  This rapid increase may be directly related to the time in which people's attention became more focused on the Iraq War in 2003. This is the central issue  that is addressed in the current study. First, we demonstrate a similar outbreak pattern using data on the spread of avian influenza A (H7N9) in China \cite{Horby2013,ZhangJ2014,Uyeki2013} along with the diffusion of disease information. Then a mathematical model was built that describes both types of spreading dynamics (i.e., disease and information) as an SIS process. Results using the model show that information diffusion can significantly inhibit epidemic spreading. Finally, simulation results exhibit multi-outbreak phenomena in the epidemic spreading accompanied by the impact of information diffusion, which is consistent with the empirical results. Our findings highlight the need for the maintenance of disease awareness even during times when epidemics appear to be under control.

\section{Results}
\subsection{Empirical Analysis}
The information diffusion during the H7N9 outbreak in China in 2013-14 was taken to illustrate the influence that information diffusion has on epidemic outbreaks. The epidemic spreading data on H7N9 infection cases was obtained from the \emph{Chinese Center for Disease Control and Prevention}. Messages that referred to "H7N9" or "avian influenza A" on \emph{Sina Weibo}, which was the largest micro-blogging system in China (\emph{http://www.weibo.com/}), were used to measure the existence of information about H7N9. The assumption was made that posted or reposted a message about H7N9 indicated an awareness of the existence of the disease; otherwise the user was considered to be unaware of it.

The spreading process of both the epidemic and information of H7N9 is illustrated in Fig.~\ref{figR1}. The blue circles ($I$) and pink diamonds ($Info$) represent the epidemic spreading and information diffusion. The similarity between the evolutionary trends of the two domains is obvious. When the epidemic broke out in Apr. 2013 and Feb. 2014 (Fig.~\ref{figR1}), evidence suggests that many people were discussing "H7N9" or "avian influenza A" in \emph{weibo.com}. This suggests that the influence of information diffusion on epidemic spreading could be quite significant. Actually, public responses to H7N9, including actions such as staying at home or wearing face masks, can affect the spread of the epidemic. Two epidemic outbreaks occurred during the period under consideration. Interestingly, the size of the first epidemic outbreak (Apr. 2013) was smaller than the second (Feb. 2014), which inversely correlated with the information domain, in which the number of individuals discussing the disease during the first outbreak was much greater than the number of individuals discussing the disease during the second outbreak. This implies that an increased awareness of H7N9 decreased the size of the epidemic outbreak. Research on news diffusion offers an alternative explanation, which suggest that as an epidemic progresses, the uncertainty surrounding it declines and the diffusion of the relevant information also decays \cite{Rosengren1973}. In the current case, it is possible that there was a higher degree of uncertainty among the Chinese people regarding H7N9 when they first heard about it in April 2013, however, they had become accustomed, and thus less anxious and responsive, to H7N9 in February 2014, even when there were more reported cases of infected people.

\subsection{Model Description}
In order to investigate the effect that information diffusion has on an epidemic outbreak, we propose a network model for the interaction between epidemic spreading and information diffusion. In this study, we consider two states of disease spreading: susceptible (S) and infected (I), and two states of information diffusion: aware (+) and unaware (-). Therefore, each individual in the system can be categorized as being in one of the four states: (\rmnum{1}) $S_{-}$: the susceptible individual who is unaware of the existence of the disease; (\rmnum{2}) $S_{+}$: the susceptible individual who is aware of the existence of the disease; (\rmnum{3}) $I_{-}$: the infected individual who is unaware of the existence of the disease; (\rmnum{4}) $I_{+}$: the infected individual who is aware of the existence of the disease.

Using the SIS model, the transformation among these states is illustrated in Fig.~\ref{figR2}. The two types of spreading processes can be described as follows:

\begin{itemize}

\item At the initial step, an individual is randomly selected and is assigned to the state $I_{+}$, which is considered as the \emph{seed} of both epidemic spreading and information diffusion. All other individuals are assigned to the $S_{-}$ state.

\item Epidemic spreading: During each time step, the infected individuals ($I_{+}$ and $I_{-}$) are capable of spreading the epidemic to their susceptible neighbors ($S_{+}$ and $S_{-}$) with the corresponding transmission probabilities and the infected individuals ($I_{+}$ and $I_{-}$) could recover to the susceptible state with the corresponding recovery probabilities.
\begin{small}
    \begin{table}[htb]
  \centering
  \begin{tabular}{p{1cm}l}
  \hline
    $\beta$                     & the probability that $S_{-}$ is infected via the $I_{-}$ neighbor ~($S_{-}I_{-} \rightarrow I_{-}I_{-}$) \\
    $\sigma_{S}\beta$           & the probability that $S_{+}$ is infected via the $I_{-}$ neighbor  ~($S_{+}I_{-} \rightarrow I_{+}I_{-}$) \\
    $\sigma_{I}\beta$           & the probability that $S_{-}$ is infected via the $I_{+}$ neighbor  ~($S_{-}I_{+} \rightarrow I_{-}I_{+}$)\\
    $\sigma_{SI}\beta$          & the probability that $S_{+}$ is infected via the $I_{+}$ neighbor  ~($S_{+}I_{+} \rightarrow I_{+}I_{+}$)\\
    $\gamma$                    & the probability that $I_{-}$ recover to $S_{-}$\\
    $\varepsilon\gamma$         & the probability that $I_{+}$ recover to $S_{+}$\\
  \hline
  \end{tabular}
\end{table}
\end{small}
\item Information diffusion: During each time step, the individuals with an awareness of the existence of the disease ($I_{+}$ and $S_{+}$) have the capability to transmit the information to their unaware neighbors ($I_{-}$ and $S_{-}$) with the probability $\alpha$. Additionally, the $I_{+}$ and $S_{+}$ individuals may become unaware of the existence of the disease with the probabilities of $\lambda$ and $\delta\lambda$ respectively.

        \begin{table}[htb]
  \centering
  \begin{tabular}{p{1cm}l}
  \hline
    $\alpha$                    & information transmission rate \\
    $\lambda$                   & information fading rate ~($S_{+}  \rightarrow S_{-} $)\\
    $\delta\lambda$             & information fading rate ~($I_{+}  \rightarrow I_{-} $)\\
  \hline
  \end{tabular}
\end{table}

\end{itemize}

The model assumes that when a susceptible individual is made aware of the existence of the disease ($S_{+}$), they will take protective measures to avoid becoming infected and $\sigma_{S}$ is denoted as the reduction in the probability of infection ($\sigma_{S}<1$). Individuals in the $I_+$ state will reduce contact with their susceptible neighbors in order to prevent the epidemic from spreading further, resulting in $\sigma_{I}<1$. With the assumption of the independent effect of the infected probability, then $\sigma_{SI}=\sigma_{S}\sigma_{I}$ when the $I_{+}$ state individuals infect the $S_{+}$ state individuals. When an $I_{+}$ individual is aware of the presence of a disease, the person will increase her/his recovery rate by taking medicine or other positive measures, which is represented by factor $\varepsilon>1$. Additionally, the $I_+$ state individuals, which could be assumed to have a better understanding of the disease, would be less likely to forget relevant disease information, leading to $\delta<1$. In the current study, since the spreading processes of information and disease are primarily determined by the corresponding transmission probabilities, we focused on the effect of the two parameters $\alpha$ and $\beta$, while fixing the other parameters. The other parameters are set as $\sigma_{S}=0.3$, $\sigma_{I}=0.6$, $\delta=0.8$, $\varepsilon=1.5$, $\lambda=0.15$ and $\gamma=0.1$, unless otherwise noted in the following analysis.

\subsection{Model Analysis}
\begin{figure*}[htb]
  \centering
  \includegraphics[width=12cm]{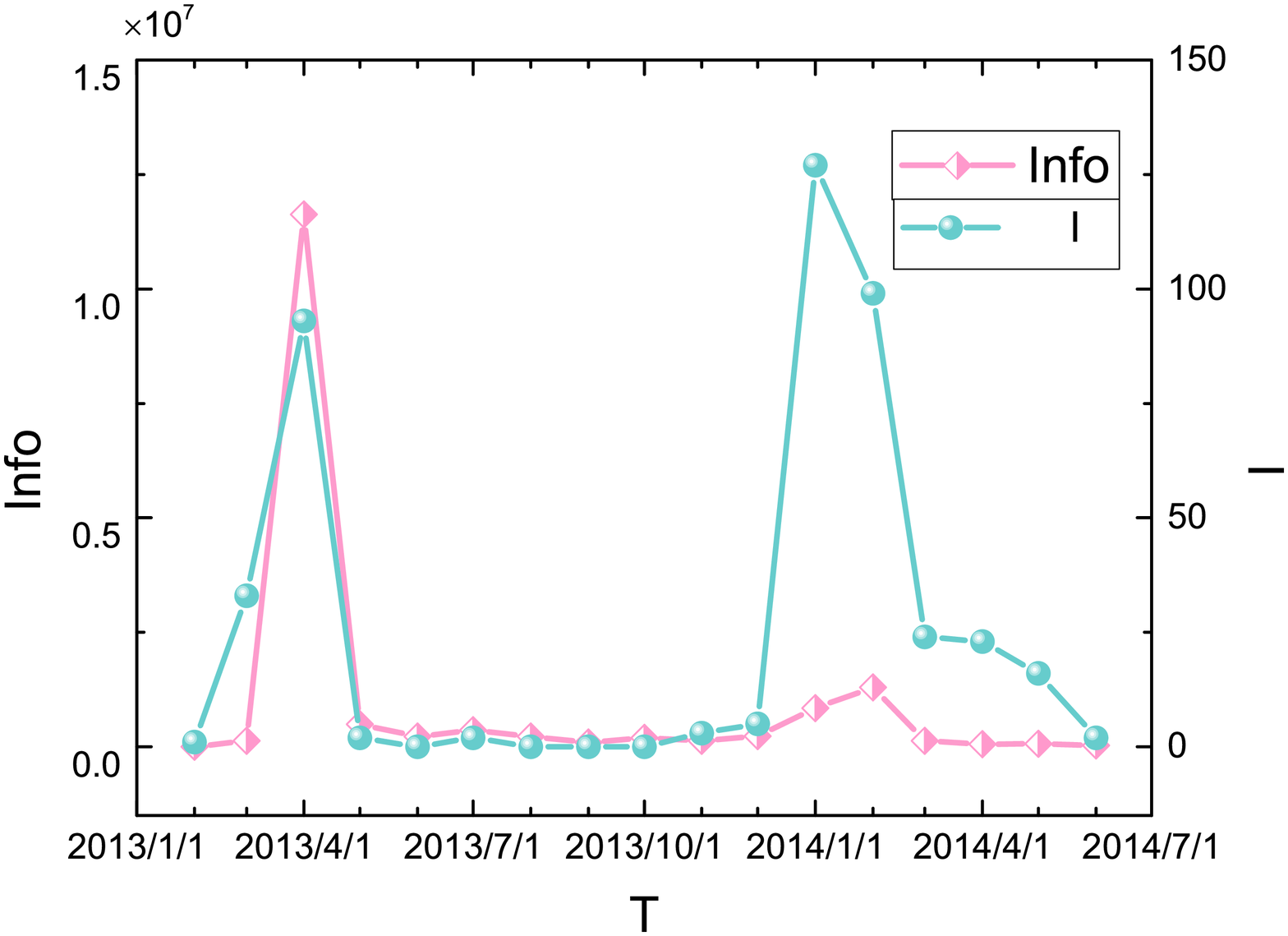}
  \caption{\label{figR1} (Color online) Empirical analysis of the epidemic spreading (blue circles) and the information diffusion (pink diamonds) of H7N9.
  }
\end{figure*}

\begin{figure*}[htb]
  \centering
  \includegraphics[width=8cm]{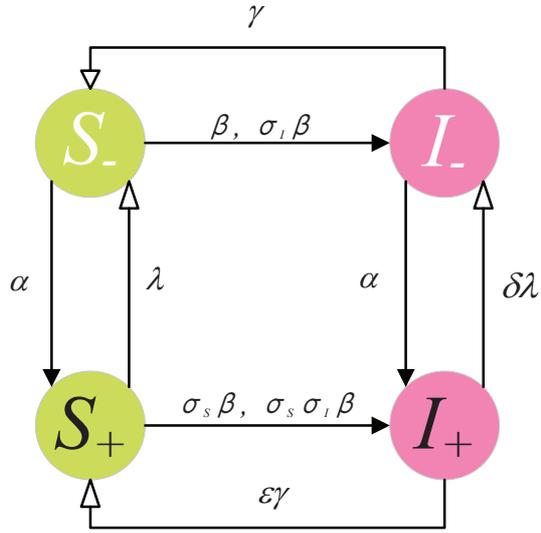}
  \caption{\label{figR2} (Color online) Diagram illustration of epidemic spreading as well as information diffusion.
  }
\end{figure*}

\begin{figure}[htb]
  \centering
  \includegraphics[width=12cm]{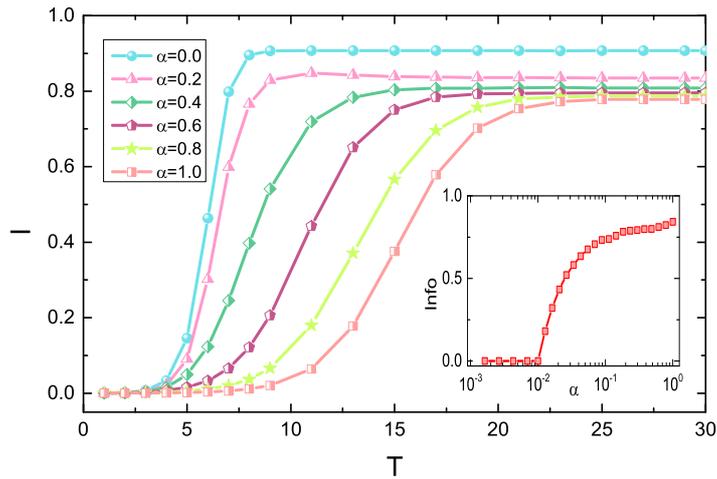}
  \caption{\label{figR3}(Color online) Dynamics of the epidemic spreading process with different $\alpha$. The inset shows the informed level as a function of $\alpha$.}
\end{figure}

The proposed model is performed on a random network with a total population $N=10000$ and an average degree $\langle k \rangle=15$.  We denote the infected level ($I$) and the informed level ($Info$) as the fraction of infected individuals (both $I_+$ and $I_-$) and the fraction of  individuals who are informed of the existence of the disease (both $S_+$ and $I_+$). The results of the simulation of the epidemic spreading process under the influence of the information diffusion when $\beta=0.3$ is shown in Fig.~\ref{figR3}. In this model, the parameter $\alpha$ can be considered as the informed level in the system, where a large $\alpha$ indicates that information spreads much easier resulting in an increase in the number of informed individuals (as in the inset of Fig. \ref{figR3}). Fig. \ref{figR3} shows that the increase in $\alpha$ leads to the decreased epidemic spreading rate and the overall diminished epidemic outbreak size. Therefore, increasing the number of informed individuals and improving self-protection measures may be an effective strategy to inhibit the spread of epidemics, which is consistent with the empirical analysis shown in Fig.~\ref{figR1}.

\begin{figure}[htb]
  \centering
  \includegraphics[width=12cm]{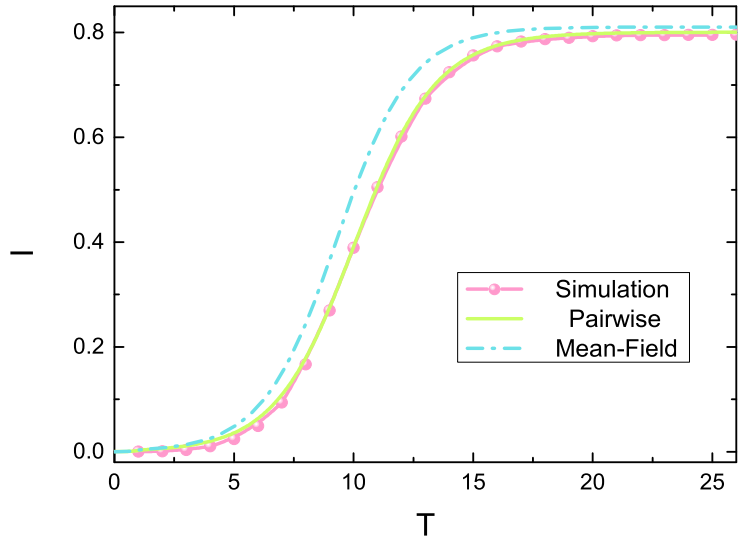}
  \caption{\label{figR4}(Color online) The mathematical analysis of the epidemic spreading process: simulation (pink circle), pairwise analysis (green solid curve) and classical mean-field analysis (blue dashed curve).}
\end{figure}

The model described in Fig.~\ref{figR2} indicates that there is a mutual feedback between information diffusion and epidemic spreading. A higher prevalence of infected individuals results in the maintenance of more informed individuals for a smaller information fading probability ($\delta<1$). This leads to the high informed level in the system, which in turn inhibits the epidemic spreading ($\sigma_{\{I,S,SI\}}<1$). This feedback effect can be clearly illustrated in the full set of differential equations based on the classical mean-field analysis\cite{Funk2009} (See Eq.(\ref{Eq:meanfield}) in Sec. {\bf{Method and Materials}}). Additionally, a full set of differential equations based on the pairwise analysis was obtained \cite{Morris1997, Keeling1999, Joo2004} that describes the interaction between the two spreading processes (See Eq.(\ref{Eq:pairwise}) in Sec. {\bf{Method and Materials}}). Actually, the pairwise analysis is also based on the mean-field assumption for calculating the change of each type of nodes as well as the node-pairs. The simulation results of the infected dynamics, numerical results of classical mean-field analysis and pairwise analysis when $\alpha=0.6$ and $\beta=0.3$ is shown in Fig.~\ref{figR4}. The results of the pairwise model are clearly a better fit to the simulation than the results of classical mean-field analysis.

\begin{figure*}[htb]
  \centering
  \includegraphics[width=15cm]{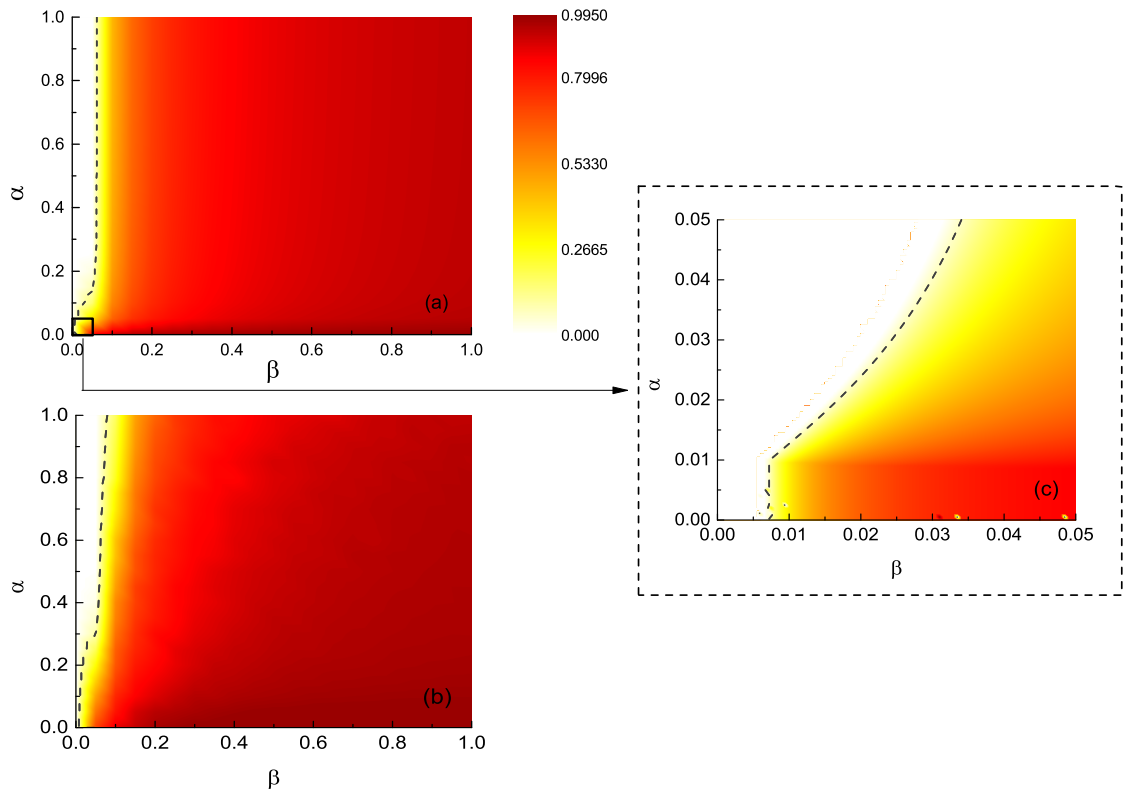}
  \caption{\label{figR5} (Color online) Comparison between pairwise analysis and simulation for the fraction of infected individuals in the stationary state (colors represent the fraction of infected individuals). (a) pairwise analysis; (b) simulation; (c) detailed view of pairwise analysis.}
 \end{figure*}

In order to investigate the effect of the mutual interaction between $\alpha$ and $\beta$ on the spreading process, we explored the full phase diagram showing the fraction of infected individuals in the whole population as a function of parameters $\alpha$ and $\beta$ in Fig.~\ref{figR5}. The results of the pairwise analysis and the simulation, which are highly consistent with each other, are shown in Fig.5 (a) and Fig.5 (b). The dashed curve in each plot displays the critical point in the epidemic spreading process, ($\beta_{c}, \alpha_{c}$), at which above this point an epidemic outbreak will occur in the population. The results clearly show that the spread of disease is much more rapid under conditions with a larger $\beta$ and smaller $\alpha$, which also indicates that the disease information diffusion can impede the the spread of disease. It should be noted that the process degenerates into a standard $SIS$ model when $\alpha=0$, the condition at which there is no information diffusion in the system, and the outbreak threshold value of the epidemic spreading is $\beta_c=\dfrac{\gamma}{\langle k \rangle}=0.0067$ \cite{Pastor-Satorras2014}. This is also consistent with the results of the pairwise analysis and the simulation shown in Fig.~\ref{figR5}. A detailed view of the pairwise analysis ($\alpha, \beta \in [0, 0.05]$ in Fig.5 (c)) was plotted in order to illustrate the threshold changes. The threshold value of $\beta$ is about 0.0067 when $\alpha<0.01$, as the disease information does not spread out in this case according to the inset of Fig. \ref{figR3}. When $\alpha>0.01$, the epidemic threshold can be significantly increased for the outbreak of the information diffusion. In other words, the information diffusion can effectively increase the epidemic threshold.

\begin{figure*}[htb]
  \centering
  \includegraphics[width=12cm]{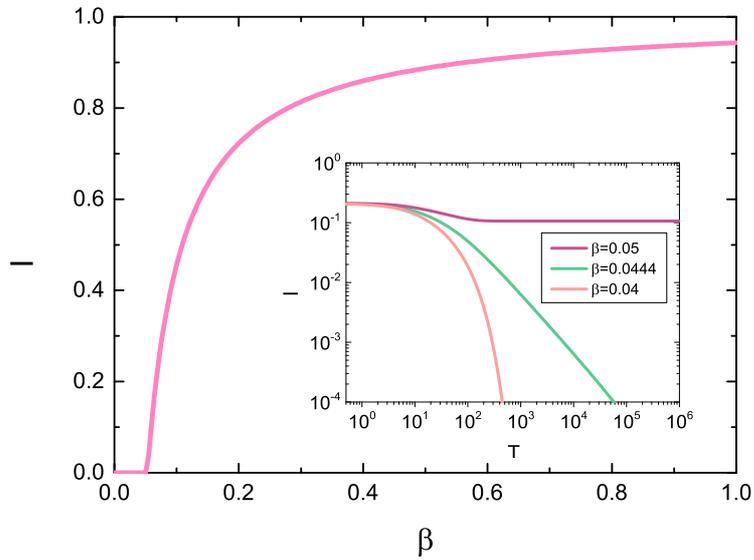}
  \caption{\label{figR6}(Color online) Infection density as a function of $\beta$ with the pairwise analysis. The inset is the infection density as a function of time with the pairwise analysis around the threshold (Different colors correspond to different $\beta$).}
\end{figure*}

The informed level is only slightly raised with the increase in $\alpha$ when $\alpha$ is large enough (e.g., $\alpha>0.3$), as shown in Fig.~\ref{figR3}, which leads to an increase in the epidemic threshold as the increase in the $\alpha$ value is not so obvious. This results also indicates that information spreading is not always an effective auxiliary measure in the control of the spread of disease, for example, in the case in which there exists a disease with a strong  infectiveness (e.g., large epidemic transmission probability $\beta$ in the red range in Fig. \ref{figR5}), enhancing the awareness of the disease alone is not enough to control the spread of disease. In order to gain a better understanding of the critical phenomenon, we investigate the evolution of the infection densities for various values of $\beta$ when $\alpha=0.6$, as shown in Fig.~\ref{figR6}.  From the differential equations of the standard SIS model, $\dfrac{dI}{dt}=-I+\beta\langle k\rangle I(1-I)$ (where $I=I_{-}+I_{+}$), we obtain $I\varpropto t^{-1}$ at the critical point with integration, which indicates that the infection density has a power-law decay along the time evolution at the critical point. The inset of Fig.~\ref{figR6} shows a power-law decay of the infection density when $\beta\thickapprox0.0444$, whereas the infected density ($I$) tends to be a steady-state value that leads to an endemic state when $\beta=0.05$, and rapidly decays to zero leading to a healthy state when $\beta=0.04$. Therefore, it can be inferred that the critical value of $\beta$ is approximately $0.0444$ in this case, which is consistent with the results shown in Fig.~\ref{figR5}, where the phase diagram indicates that the critical value of $\beta$ is around $0.045$ when $\alpha=0.6$.

Interestingly, the empirical analysis illustrates that the dynamics of many diseases exhibit a multi-outbreak phenomena \cite{Zhang2011,Fang2013,Wang2014,Kan2008}, in which there are several waves in the process of epidemic spreading, similar to the dynamics illustrated in Fig.~\ref{figR1}. Generally,  there are many complex factors that contribute to multi-outbreaks, including seasonal influence, climatic variation, and incubation period. The periodic outbreaks in the SIS model can be interpreted by the influence of information diffusion. As previously mentioned, there is a mutual feedback between information diffusion and epidemic spreading in the proposed model. On the one hand, a larger proportion of infected individuals should result in an increase in preventive behavioral responses \cite{Sahneh2012} due to the increased awareness of the disease, thus leading to a steady decrease in infected cases over time. On the other hand, when the spread of the disease appears to be under control (e.g., the size of infected population decreases), people become less vigilant, which leads to a decrease in the dissemination of information and increases the chances of a second outbreak. Notably, the size of the second outbreak is often larger than the previous one, as in the case of SARS in 2003 (Fig.2 in Ref. \cite{WorldHO2003}), \emph{Dengue Fever} in Taiwan in 2001-2002 (Fig.3A in Ref. \cite{Kan2008}).

\begin{figure*}[htb]
  \centering
  \includegraphics[width=12cm]{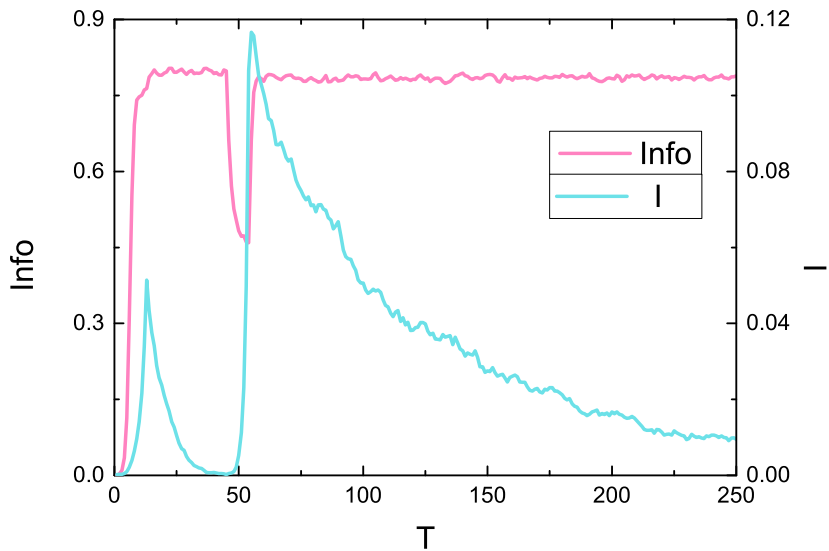}
  \caption{\label{figR7} (Color online) Multi-outbreak phenomena of the epidemic spreading with the influence of the information diffusion. $I_{high}$ and $I_{low}$ are set as 0.05 and 0.0003 respectively.
  }
\end{figure*}

In order to illustrate the multi-outbreak phenomena of epidemic spreading with the influence of information diffusion, two critical infected levels were set in this model for simplicity  ($I_{high}$ and $I_{low}$). When the fraction of infected individuals in the population is larger than the critical value of the high infected level ($>I_{high}$), the information spreads more quickly and there is an increase in preventative behavioral responses. Based on this assumption, we obtain simulation results of the dynamic of infected individuals, as shown in Fig.~\ref{figR7}. In this case, we set $\beta=0.18$, which is much larger than the epidemic threshold without considering information diffusion. The disease spreads very quickly at first while there are very few people who are aware of the occurrence of the disease. Typically, the diffusion of information regarding a disease occurs at a faster rate than epidemic spreading. Therefore, as the information regarding the disease quickly spreads, the high informed level has a significant effect on inhibiting the spread of the epidemic (the decay period of the epidemic). It should be noted that the epidemic will be completely suppressed if the high informed level can be maintained. However, when the disease is controlled from the first outbreak (i.e., the infected density is smaller than $I_{low}$), members of the population are likely to no longer consider the disease as a threat, thus ignoring the disease propagation and no longer actively engaging in protective measures, which will in turn lead to another outbreak of the epidemic,  with a high probability of the second being larger than the first. In this model, such changes in behavioral responses can be illustrated by using different parameter settings, such as small $\alpha$ and large $\sigma_I$ and $\sigma_S$. This allows for the disease to spread again as long as social awareness becomes low enough, as is shown in the second epidemic outbreak illustrated in Fig. ~\ref{figR7}. Similar to the empirical analysis shown in Fig. ~\ref{figR1}, the size of the first epidemic outbreak is smaller than that of the second one, whereas the informed level in the first epidemic outbreak is higher than that in the second one. It should be noted that, due to the difficulty in precisely quantifying the informed level in the empirical analysis, the number of tweets that discuss the related disease is used as a proxy measure in Fig. \ref{figR1}. In contrast to the trend shown in Fig. \ref{figR1}, the high informed level must be maintained during the period when the infected proportion decreases as shown by the pink curve in Fig. \ref{figR7}. Based on the model analysis, it could be concluded that it is important to strengthen public awareness of disease occurrence, especially during times in which the spread of the epidemic is under control, otherwise, there is a high probability of a second outbreak.

\section{Conclusions \& Discussion}
\label{Sec:Conclusion}
In this study, we have studied the interaction between epidemic spreading and relevant information diffusion. The empirical analysis shows that information diffusion can significantly inhibit epidemic spreading, in which the size of the epidemic outbreak is influenced by the informed level. In line with previous works on the closed feed-back loop ''epidemic spreading $\rightarrow$ behavior change $\rightarrow$ epidemic spreading" \cite{Funk2009,ZhangHF2014}, we build a model in which the two types of spreading dynamics are described as an SIS process. Both the simulation results and the pairwise analysis reveal that information diffusion can increase the epidemic outbreak threshold, diminish the final fractions of infected individuals and significantly slow down the rate of propagation. More importantly, we address the issue of multi-outbreak phenomena of epidemic spreading with information diffusion as the governing. The results of the simulation suggest that a higher epidemic prevalence impels people to increase in disease information sharing, leading to a high level of informed individuals, which in turn results in the steady decrease in the number of infected cases. During the periods in which the disease appears to be controlled, less attention is given to the disease leading to a decrease in the transmission of information and an increase in the chance of another outbreak. Additionally, the simulation results of the multi-outbreak phenomena were consistent with the empirical analysis.

The findings from this work support the idea that preventive behavioral measures brought about by disease information can significantly inhibit the spread of an epidemic, and the increase in information diffusion can be utilized as an auxiliary measure to efficiently control epidemics. The government should make an effort to maintain social awareness of the disease, even during times in which the epidemic seems to be under control, in order to prevent another outbreak. In this study, we focus on the inhibition of information diffusion, and preventive behavioral responses are illustrated with some parameters generally. However, the dynamics of an epidemic may be very different depending on the behavioral responses of people, such as adaptive process \cite{Gross2006}, migration \cite{Cui2011}, vaccination \cite{ZhangHF2014}, and immunity \cite{Pulliam2007}. This work provides a basic understanding of the interplay between the two spreading processes. Future research should focus on the in-depth study of preventive behavioral responses induced by diffusion of disease information.

\section{Methods and Materials}

In this study, $[\ast]$ represents the number of state variables ($\ast$) in the system at time step $t$. $[S_+]$, $[S_-]$, $[I_+]$ and $[I_-]$ represent the number of aware susceptible, unaware susceptible, aware infected and unaware infected. In the pairwise analysis, $[\ast]$ also denotes the number of the corresponding state variable of the edges, for example, $[S_+I_+]$ represents the number of edges between two individuals at states $S_{+}$ and $I_{+}$.

\noindent{\bf{Mean-field Analysis:}} According to Fig. \ref{figR2}, we adopt mean-field analysis for the spread of epidemic and information in a homogeneous network as follows:
\begin{equation}\footnotesize
 \label{Eq:meanfield}
 \left\{
  \begin{split}
  \frac{d[S_{-}]}{dt}=&-\langle k \rangle\beta [I_{-}]\frac{[S_{-}]}{N}-\langle k \rangle\sigma_{I}\beta [I_{+}]\frac{[S_{-}]}{N}-\langle k \rangle\alpha([S_{+}]+[I_{+}])\frac{[S_{-}]}{N}+\lambda [S_{+}]+\gamma [I_{-}]\\[5pt]
  \frac{d[S_{+}]}{dt}=&-\langle k \rangle\sigma_{S}\beta [I_{-}]\frac{[S_{+}]}{N}-\langle k \rangle\sigma_{S}\sigma_{I}\beta [I_{+}]\frac{[S_{+}]}{N}+\langle k \rangle\alpha([S_{+}]+[I_{+}])\frac{[S_{-}]}{N}-\lambda [S_{+}]+\varepsilon\gamma [I_{+}]\\[5pt]
  \frac{d[I_{-}]}{dt}=&\langle k \rangle\beta [I_{-}]\frac{[S_{-}]}{N}+\langle k \rangle\sigma_{I}\beta [I_{+}]\frac{[S_{-}]}{N}-\langle k \rangle\alpha([S_{+}]+[I_{+}])\frac{[I_{-}]}{N}+\delta\lambda [I_{+}]-\gamma [I_-]\\[5pt]
  \frac{d[I_{+}]}{dt}=&\langle k \rangle\sigma_{S}\beta [I_{-}]\frac{[S_{+}]}{N}+\langle k \rangle\sigma_{S}\sigma_{I}\beta [I_{+}] \frac{[S_{+}]}{N}+\langle k \rangle\alpha([S_{+}]+[I_{+}])\frac{[I_{-}]}{N} -\delta\lambda [I_{+}]-\varepsilon\gamma [I_{+}]\\
   \end{split}
   \right.
\end{equation}
Where $N$ is the number of individuals in the system, $\langle k \rangle$ is the average degree of the network and the other parameters are illustrated in \textbf{Nomenclature}.

\noindent{\bf{Pairwise Analysis:}} Pairwise models have recently been widely used to illustrate the dynamic process of epidemics on networks, as those models take into account the edges of the networks \cite{Morris1997, Keeling1999, Joo2004}. In this study, we consider a set of evolution equations which are comprised of four types of individuals and 10 types of edges. Using the well-known closure, expressed as $[ABC]=\dfrac{[AB][BC]}{[B]}$ (assuming the neighbors of each individual obey Poisson distribution)~\cite{Morris1997}, we can get a set of differential equations as follows :
\begin{equation}\footnotesize
\label{Eq:pairwise}
 \left\{
 \begin{split}
 \frac{d[S_{-}]}{dt}=&-\beta[S_{-}I_{-}]-\sigma_{I}\beta[S_{-}I_{+}]-\alpha([S_{-}S_{+}]+[S_{-}I_{+}])+\lambda[S_{+}]+\gamma[I_{-}]\\[5pt]
 \frac{d[S_{+}]}{dt}=&-\sigma_{S}\beta[S_{+}I_{-}]-\sigma_{S}\sigma_{I}\beta[S_{+}I_{+}]+\alpha([S_{-}S_{+}]+[S_{-}I_{+}])-\lambda[S_{+}]+\varepsilon\gamma[I_{+}]\\[5pt]
 \frac{d[I_{-}]}{dt}=&\beta[S_{-}I_{-}]+\sigma_{I}\beta[S_{-}I_{+}]-\alpha([S_{+}I_{-}]+[I_{-}I_{+}])+\delta\lambda[I_{+}]-\gamma[I_{-}]\\[5pt]
 \frac{d[I_{+}]}{dt}=&\sigma_{S}\beta[S_{+}I_{-}]+\sigma_{S}\sigma_{I}\beta[S_{+}I_{+}]+\alpha([S_{+}I_{-}]+[I_{-}I_{+}])-\delta\lambda[I_{+}]-\varepsilon\gamma[I_{+}]\\[5pt]
 \frac{d[S_{-}I_{-}]}{dt}&=-\beta[S_{-}I_{-}]+\lambda([S_{+}I_{-}]+\delta[S_{-}I_{+}])+\beta\frac{[S_{-}I_{-}]([S_{-}S_{-}]-[S_{-}I_{-}])}{[S_{-}]}+\sigma_{I}\beta\frac{[S_{-}I_{+}]([S_{-}S_{-}]-[S_{-}I_{-}])}{[S_{-}]}\\
                          &-\alpha\frac{[S_{-}I_{-}]([S_{-}I_{+}]+[S_{-}S_{+}])}{[S_{-}]}-\alpha\frac{[S_{-}I_{-}]([I_{-}I_{+}]+[S_{+}I_{-}])}{[I_{-}]}-\gamma[S_{-}I_{-}]+\gamma[I_{-}I_{-}]\\[5pt]
 \frac{d[S_{-}I_{+}]}{dt}&=-\sigma_{I}\beta[S_{-}I_{+}]+\lambda[S_{+}I_{+}]-\alpha[S_{-}I_{+}]-\delta\lambda[S_{-}I_{+}]-\beta\frac{[S_{-}I_{-}][S_{-}I_{+}]}{[S_{-}]}+\sigma_{S}\beta\frac{[S_{+}I_{-}][S_{-}S_{+}]}{[S_{+}]}-\sigma_{I}\beta\dfrac{[S_{-}I_{+}]^2}{[S_{-}]}\\
                          &+\sigma_{S}\sigma_{I}\beta\frac{[S_{+}I_{+}][S_{-}S_{+}]}{[S_{+}]}+\alpha\frac{[S_{-}I_{-}]([I_{-}I_{+}]+[S_{+}I_{-}])}{[I_{-}]}-\alpha\frac{[S_{-}I_{+}]([S_{-}I_{+}]+[S_{-}S_{+}])}{[S_{-}]}-\varepsilon\gamma[S_{-}I_{+}]+\gamma[I_{-}I_{+}]\\[5pt]
 \frac{d[S_{+}I_{-}]}{dt}&=-\sigma_{S}\beta[S_{+}I_{-}]+\delta\lambda[S_{+}I_{+}]-\lambda[S_{+}I_{-}]-\alpha[S_{+}I_{-}]-\sigma_{S}\beta\frac{[S_{+}I_{-}]^{2}}{[S_{+}]}-\sigma_{S}\sigma_{I}\beta\frac{[S_{+}I_{+}][S_{+}I_{-}]}{[S_{+}]}+\beta\frac{[S_{-}I_{-}][S_{-}S_{+}]}{[S_{-}]}\\
                          &+\sigma_{I}\beta\frac{[S_{-}I_{+}][S_{-}S_{+}]}{[S_{-}]}+\alpha\frac{[S_{-}I_{-}]([S_{-}S_{+}]+[S_{-}I_{+}])}{[S_{-}]}-\alpha\frac{[S_{+}I_{-}]([I_{-}I_{+}]+[S_{+}I_{-}])}{[I_{-}]}-\gamma[S_{+}I_{-}]+\varepsilon\gamma[I_{-}I_{+}]\\[5pt]
 \frac{d[S_{+}I_{+}]}{dt}&=-\sigma_{S}\sigma_{I}\beta[S_{+}I_{+}]+\alpha[S_{-}I_{+}]+\alpha[S_{+}I_{-}]+\sigma_{S}\beta\frac{[S_{+}I_{-}]([S_{+}S_{+}]-[S_{+}I_{+}])}{[S_{+}]}+\sigma_{S}\sigma_{I}\beta\frac{[S_{+}I_{+}]([S_{+}S_{+}]-[S_{+}I_{+}])}{[S_{+}]}\\
                          &+\alpha\frac{[S_{-}I_{+}]([S_{-}I_{+}]+[S_{-}S_{+}])}{[S_{-}]}+\alpha\frac{[S_{+}I_{-}]([S_{+}I_{-}]+[I_{-}I_{+}])}{[I_{-}]}-\varepsilon\gamma[S_{+}I_{+}]+\varepsilon\gamma[I_{+}I_{+}]-\lambda[S_{+}I_{+}]-\delta\lambda[S_{+}I_{+}]\\[5pt]
 \frac{d[I_{-}I_{-}]}{dt}&=2\beta[S_{-}I_{-}]+2\delta\lambda[I_{-}I_{+}]]+2\beta\frac{[S_{-}I_{-}]^{2}}{[S-]}+2\sigma_{I}\beta\frac{[S_{-}I_{+}][S_{-}I_{-}]}{[S_{-}]}-2\alpha\frac{[I_{-}I_{-}]([S_{+}I_{-}]+[I_{-}I_{+}])}{[I_{-}]}-2\gamma[I_{-}I_{-}]\\[5pt]
 \frac{d[I_{-}I_{+}]}{dt}&=\sigma_{I}\beta[S_{-}I_{+}]+\sigma_{S}\beta[S_{+}I_{-}]+\delta\lambda([I_{+}I_{+}]-[I_{-}I_{+}])-\alpha[I_{-}I_{+}]+\beta\frac{[S_{-}I_{-}][S_{-}I_{+}]}{[S_{-}]}+\sigma_{I}\beta\frac{[S_{-}I_{+}]^{2}}{[S_{-}]}+\sigma_{S}\beta\frac{[S_{+}I_{-}]^{2}}{[S_{+}]}\\
                          &+\sigma_{S}\sigma_{I}\beta\frac{[S_{+}I_{+}][S_{+}I_{-}]}{[S_{+}]}+\alpha\frac{[I_{-}I_{-}]([S_{+}I_{-}]+[I_{-}I_{+}])}{[I_{-}]}-\alpha\frac{[I_{-}I_{+}]([S_{+}I_{-}]+[I_{-}I_{+}])}{[I_{-}]}-\varepsilon\gamma[I_{-}I_{+}]-\gamma[I_{-}I_{+}]\\[5pt]
 \frac{d[I_{+}I_{+}]}{dt}&=2\sigma_{S}\sigma_{I}\beta[S_{+}I_{+}]+2\alpha[I_{-}I_{+}]-2\delta\lambda[I_{+}I_{+}]+2\sigma_{S}\beta\frac{[S_{+}I_{-}][S_{+}I_{+}]}{[S_{+}]}+2\sigma_{S}\sigma_{I}\beta\frac{[S_{+}I_{+}]^{2}}{[S_{+}]}\\
                         &+2\alpha\frac{[I_{-}I_{+}]([S_{+}I_{-}]+[I_{-}I_{+}])}{[I_{-}]}-2\varepsilon\gamma[I_{+}I_{+}]\\[5pt]
 \frac{d[S_{-}S_{-}]}{dt}&=2\lambda[S_{-}S_{+}]-2\beta\frac{[S_{-}I_{-}][S_{-}S_{-}]}{[S_{-}]}-2\sigma_{I}\beta\frac{[S_{-}I_{+}][S_{-}S_{-}]}{[S_{-}]}-2\alpha\frac{[S_{-}S_{-}]([S_{-}S_{+}]+[S_{-}I_{+}])}{[S_{-}]}+2\gamma[S_{-}I_{-}]\\[5pt]  \frac{d[S_{-}S_{+}]}{dt}&=\lambda[S_{+}S_{+}]-\lambda[S_{-}S_{+}]-\alpha[S_{-}S_{+}]-\sigma_{S}\beta\frac{[S_{+}I_{-}][S_{-}S_{+}]}{[S_{+}]}-\sigma_{S}\sigma_{I}\beta\frac{[S_{+}I_{+}][S_{-}S_{+}]}{[S_{+}]}-\beta\frac{[S_{-}I_{-}][S_{-}S_{+}]}{[S_{-}]}\\
                          &-\sigma_{I}\beta\frac{[S_{-}I_{+}][S_{-}S_{+}]}{[S_{-}]}+\alpha\frac{[S_{-}S_{-}]([S_{-}S_{+}]+[S_{-}I_{+}])}{[S_{-}]}-\alpha\frac{[S_{-}S_{+}]([S_{-}S_{+}]+[S_{-}I_{+}])}{[S_{-}]}+\varepsilon\gamma[S_{-}I_{+}]+\gamma[S_{+}I_{-}]\\[5pt]
 \frac{d[S_{+}S_{+}]}{dt}&=2\alpha[S_{-}S_{+}]-2\lambda[S_{+}S_{+}]-2\sigma_{S}\beta\frac{[S_{+}I_{-}][S_{+}S_{+}]}{[S_{+}]}-2\sigma_{S}\sigma_{I}\beta\frac{[S_{+}I_{+}][S_{+}S_{+}]}{[S_{+}]}\\
                         &+2\alpha\frac{[S_{-}S_{+}]([S_{-}S_{+}]+[S_{-}I_{+}])}{[S_{-}]}-2\varepsilon\gamma[S_{+}I_{+}]
 \end{split}
 \right.
\end{equation}

\section*{Acknowledgments}

\noindent
{\bf Funding statement.} This work was partially supported by Natural Science Foundation of China (Grant Nos. 11305043 and 11301490), Zhejiang Provincial Natural Science Foundation of China (Grant No. LY14A050001), Zhejiang Qianjiang Talents Project (QJC1302001), the EU FP7 Grant 611272 (project GROWTHCOM) and Hong Kong Research Grants Council GRF (CityU 154412).



\clearpage

\end{document}